# Electrochemical and *in-situ* synchrotron studies of Pt-Ru-Pb ternary catalyst for direct methanol fuel cell

or

# Evaluating the role of Lead In A Novel Ternary Catalysts For DMFC s

**Aditi Halder, Qingying Jia, Matthew Trahan and Sanjeev Mukerjee**

**Department of Chemistry and Chemical Biology**

**Northeastern University,360 Huntington Avenue,Boston,MA 02115,USA**

Corresponding author:

Email:s.mukerjee@neu.edu

Tel.6173731282

Fax.6173732894

**Abstract**

The current density at lower potential is highly desirable in fuel cell technology and crucial center point for designing a new catalyst. By alloying platinum with various other metals, the improvisation of the fuel cell catalyst has achieved a lot of attention and interests. In this article, a novel porous ternary alloy PtPb@Ru as anode catalysts for direct methanol fuel cell had been synthesized by micro-emulsion technique. The catalysts had been characterized by various spectroscopic and microscopic techniques. The activity and durability of the catalysts had been tested by running cyclic voltammetry in 0.1 M $HClO_4$ and 1M Methanol. To explain the many fold increase in current density of the PtPb@Ru catalysts in comparison to the commercial available PtRu catalysts, in situ X-ray absorption spectroscopy (XAS) measurements, at the $PtL_3$ edge (XANES and EXAFS) were carried out on the PtPb@Ru catalysts in an electrochemical cell. The down-shift in the d-band center of platinum observed in the XAS study, might be responsible for the better activity and high current density observed here.

**Introduction:**

The gradual depletion in natural resources and the increasing energy demand make energy as ""the single most important scientific and technological challenge facing humanity in the next century.[1]" The energy solution for the future generation definitely cannot be dependent upon the natural resources due to the gradual diminution in the fossil fuel reservoirs. The increase in environmental pollution and increment in the global energy consumption, led the slow but steady development of alternative and renewable source of energy. The direct methanol fuel cell is one of these sources of energy which has drawn a huge amount of attention because of its high energy density, efficiency and low green-house gas emission. The advantage of direct methanol fuel cell over hydrogen fuel cell, especially for portable power application, is because of its easy transportability, safety storage and low temperature operating conditions[2]. At the structural perspective, DMFC is compatible with the existing petroleum distribution network, and no necessity of fuel reformer, complex humidification and heat management.

The performance of the direct methanol fuel cell largely depends on two factors, (1) membrane used in the membrane electrode assembly (MEA) because of methanol crossover and cross-membrane contamination of the catalysts; and (2) the quality of the electrocatalysts used in the anode and cathode. The choice of the anode electrocatalysts is very crucial for DMFC as the reaction kinetics, current density and cell durability largely depends upon the quality of the catalysts. Undoubtedly, the best anode catalyst for DMFC is platinum which shows the best electrocatalytic activity and stability in acidic environment. However, platinum has its own disadvantage; the strong affinity of platinum towards carbon monoxide (the reaction intermediate of the methanol oxidation) is one of the major reasons for the poor performance of the DMFC. Due to the CO poisoning of the anode catalysts, the available surface sites on the platinum reduces and ultimately decreases the overall cell performance. To mitigate the CO poisoning into anode electrode and enhance the available sites of more methanol adsorption on platinum surfaces, the use of second metal e.g. Ru is well established technique, known as “bi-functional mechanism”[3]. In bi-functional mechanism, the oxophilic nature of Ru has been exploited for the generation of the hydroxyl groups which aids the oxidative removal of CO bound on the platinum surfaces. Several other research works had been done on the optimization of the activity of catalysts with other metals[2, 4-16] towards methanol oxidation and however, PtRu bimetallic catalyst is known to be the best so far. Unfortunately, certain issues related to the stability and durability of the state-of-the art Pt-Ru catalysts had led new research area for improving/synthesizing a new catalyst. The poor stability of PtRu alloys surfaces due to the low energy of formation[5] and segregation of Ru[17] on surfaces make the catalysts vulnerable and poor in durability. The result is the dissolution of ruthenium from the PtRu catalysts under fuel cell operating condition and hydro-osmotic dragging of ruthenium through the polymer membrane towards the cathodic part of the fuel cell, decreasing the overall performance of the fuel cell[18-20].

In this current scenario, a number of other catalysts have been synthesized, and development of ternary catalysts has drawn a quite attention. Here in this paper, we are reporting a new ternary catalyst which is based on the property of PtRu, but having a third metal on the outer shell. A number of studies have reported where Pb enhances the activity of Pt[5, 21] , some cases it has been suggested that lead has a promoting effect on the oxidation of alcohol and small organic molecules[22]. The large difference in the work function of lead which is 4.75eV and platinum (which is 5.12-5.39 eV) leads to the probability of the charge transfer from lead to platinum surfaces. Earlier report of lead deposited on the platinum (111) surfaces by under potential technique had been analyzed to investigate the influence of lead on the catalytic activity of platinum[23] and scanning tunneling microscopy(STM) study showed that there is a strong interaction between the Pb and Pt atoms due to their large difference in the work function[24]. The 20 times increase in reaction rate for formic acid oxidation by under potentially deposited lead[23] had been also reported. The lead-based ternary catalyst which we are reporting in this article also shows very high current density and good stability. To understand the higher activity of this ternary catalyst and to evaluate the exact role of lead in the ternary catalyst; along with electrochemical study, the in situ X-ray Absorption Spectroscopy had been done under fuel cell operating condition. The new results led us to conclude a new role played by Pb which is very intersecting and explained the previous data collected by other groups and theoretical study reported elsewhere[25]. This study also leads to the idea that using lead(Pb) in a controlled amount in a ternary catalysts for DMFC application is indeed an useful one.

## Experimental Section

### 1. Synthesis method of Ternary Catalysts

The lead-based core-shell ternary catalysts have been made by a sequential two step microemulsion techniques. All syntheses of carbon-supported PtRu@Pb catalysts were carried out in an organic atmosphere using a Schlenk technique. Sodium dioctyl sufosuccinate (AOT) was used as surfactant and reverse micelle solutions were made up of an organic phase cyclohexane, surfactant AOT.

### 1.1 Synthesis of PtRu@Pb

In the first step a surfactant solution of AOT was made in cyclohexane and in that a mixture solution of aqueous solution of 1.01 ml of $H_2PtCl_6$ (0.25M) and 1.69ml of 0.15M of $RuCl_3.3H_2O$ was added. The reverse micelle solution was stirred and after stirring for 1 hour, another reverse micelle solution of reducing agent sodium borohydride was slowly added into this. The solution was slowly turned black, ensuring the formation of platinum-ruthenium alloy nanoparticles. To prepare the core-shell nanoparticle, a reverse micelle solution containing 127 µl of $Pb(NO_3)_2$(0.15M) was slowly added into the platinum-ruthenium alloy solution. On reduction by sodium borohydride, lead (Pb) metal was nucleated on the surface of the core platinum-ruthenium alloys. The complete reduction of lead resulted in the formation of platinum-ruthenium core and outer layer shell of lead metal.

After synthesizing the core-shell nanoparticles, the nanoparticles should be impregnated into carbon support. The carbon (Ketjan Black) was heat-treated for 110°C for 6 hours under vacuum and dispersed into 50ml of cyclohexane. The carbon solution then slowly added into the metal alloy solution and stirred for overnight (12hours) under argon. A 180 ml solution of acetone was slowly added into the reverse micelle mixture and controlled addition of acetone was required to prevent the rapid breakage of micelles to avoid agglomeration of nanoparticles. The solution kept undisturbed for at least 2hours. The carbon-supported metal nanoparticles then washed with acetone, ethanol and Millipore water and filtered using sub-micrometer-pore-sized filter (0.22µm Millipore filter). The product was dried under vacuum for 8 hours at 70°C.

The carbon-supported core-shell nanoparticles were heat-treated under hydrogen (5%)-Argon atmosphere for 6 hours at a ramp rate of 5°C $min^{-1}$ to the desired temperature 220°C .

## 2. Physical Characterizations

The carbon-supported core-shell nanoparticles had been characterized for X-ray diffraction (XRD), X-ray photoelectron spectroscopy (XPS), energy dispersive X-ray spectroscopy (EDS) and transmission electron microscope(TEM) analyses. Powder X-ray diffraction (XRD) patterns of the catalysts had been collected by using a Rigaku X-ray Diffractometer equipment with CuKα( wavelength 1.5418Å) filtered by a nickel filter. The samples were scanned from 10°<2θ<90° and the recorded patterns were matched against the powdered patterns using the JCPDS files. The morphology and compositional analysis of the samples had been done using a JEOL 2100 transmission electron microscope using an energy dispersive X-ray spectroscopy (EDS) analyses. The X-ray photoelectron spectroscopy had been carried out in XPS ESCA SSX-100 machine to understand the surface nature, using Al-$K_{\alpha}$ radiation source. To obtain the XPS spectra, the pressure of the analysis chamber was maintained at $10^{-10}$ torr. The binding energy (BE) and the kinetic energy (KE) scales were adjusted by setting the C1*s* transition at 284.6 eV.

**X-ray absorption spectroscopy (XAS) measurements & analysis**

All experiments were carried out in an *in-situ* specially designed electrochemical XAS half cell at room temperature at beamline X-3B at the National Synchrotron Light Source (NSLS), Brookhaven National Lab. The cell design has been reported in previous work[20]. The working electrode in the cell was one piece of 2 $cm^2$ carbon cloth painted with PtRu@Pb catalysts. A carbon cloth was used as the counter electrode. A sealed reversible hydrogen electrode was used as the reference electrode. De-aerated 0.1 M $HClO_4$ solution was used as the electrolyte.

Cyclic Voltammetry was taken between the potential windows of 0.02V-1V. After achieving stable CVs, full range Pt $L_3$ edge extended X-ray absorption fine structure (EXAFS) scans were collected at various static potentials along the anodic sweep of the CV in fluorescence mode with Pt foil as the reference for accurate energy calibration. Afterward, 1 ml methanol was dripped into the electrolyte to reach the concentration of 1 M, and then the procedure above was repeated. EXAFS data analysis was carried out using the IFEFFIT suite[26] (version 1.2.9,

IFEFFIT Copyright 2005, Matthew Newville, University of Chicago). The $\chi(R)$ were modeled using single scattering paths calculated by FEFF8[27].

The $\Delta\mu$ analysis technique has been described in great detail elsewhere.[28, 29] Briefly, FEFF8 calculations were performed to interpret the $\Delta\mu$ spectra using $Pt_4M_2$ cluster as shown in figure 5(a) . In all of the calculations, unless otherwise noted, the theoretical $\Delta\mu$ is calculated as:

$$\Delta\mu = \mu(ads / Pt_4Ru_2) - \mu(Pt_4Ru_2) \quad (1)$$

where the $\mu(Pt_4Ru_2)$ is obtained from the clean cluster $Pt_4Ru_2$ with Pt–Pt bond distances of 2.72 Å and Pt–Ru distances of 2.67 Å as guided by the experimental EXAFS results, ads stands for adsorbates such as CO, O(H) and Pb (Pb is treated as an adsorbate here since it forms a outer layer shell on top of the PtRu core). The Pt-O(H) and Pt-CO bond distance is set as 2.0 Å.

## Results and Discussion

### Structural and Microstructural Studies

The XRD of ternary PtRu@Pb catalysts had been analyzed and compared with the commercial PtRu-ETEK catalysts (Figure.1A). The formation of crystalline and alloyed phases of PtRu had been confirmed by the reflection from the (111), (200), (220) and (311) planes. The positive shift in the PtRu@Pb catalysts in comparison to the PtRu(commercial) catalysts indicated the incorporation of lead in the crystal lattice, though the reflection of any lead or lead oxide was absent because of very low concentration of lead present in the catalysts. The broad XRD peak of PtRu@Pb catalysts depicted the nano-crystalline nature of the sample and an estimation of the particle size using Scherrer's formula gives the average crystallite size of 3-5nm.

The morphology of the ternary PtRu@Pb had been studied by high resolution transmission electroscopy (HRTEM) (Figure 2.a) and it revealed that this ternary catalyst was actually comprising of porous clusters of 50 nm in size (Figure 2.b), supported on carbon support. Each individual nanocluster was again made up of 3-5nm nanoparticles (Figure 2.c) and high resolution TEM showed the lattice fringes of the alloys and moire-fringes due to overlapping of nanocrystals of different lattice parameters(figure d). That clearly confirmed that some of lead nanoparticles did not alloy with PtRu but instead it formed a mixture solution. The energy dispersive X-ray spectroscopy (EDS) analyses from a single porous cluster gave a compositional analysis from a single cluster which showed that the weight ratio of platinum, ruthenium and lead was 72:25:2.7. The EDS analysis confirmed the very low concentration of lead in the catalysts sample (Figure 1B).

After completing the structural and micro-structural characterization, the electrochemical analyses were performed to study the electrocatalytic activities of all the catalysts, 0.1 M $HClO_4$ solution was used as the earlier reports[30], that the lead ad-atoms showed an inhibiting effect on the oxidation of alcohol in sulfuric acid. Thus, to avoid the adverse effect of sulfuric acid, all the electrochemical studies have been performed in perchloric acid. Figure 3 shows the comparison in the onset of oxide formation for a thin film of 60% PtRu –ETEK catalysts, PtRu@PbRT (without heat treated) and PtRu@Pb 220(heat treated at 220°C) catalysts in Ar-saturated 0.1 M $HClO_4$ solution after cyclicing the electrodes for several times until steady voltammogram obtained. The loading of the catalysts material on the 5 mm diameter glassy-carbon electrode was 15μg/$cm^2$. The scan rate was 10mV/sec. The cyclic voltammograms of the commercial PtRu-ETEK (60%) and both PtRu@PbRT and PtRu@Pb220 catalysts had distinct differences especially in the hydrogen adsorption/desorption region. The carbon supported commercial PtRu-ETEK catalyst possess some degree of low –coordinated crystalline planes[31] and hence the hydrogen adsorption-desorption features between 0.4 and 0 V vs RHE are different than the

PtRu@Pb catalysts. In the case of commercial PtRu-ETEK (60%) a distinct oxide formation peak at around 0.7 V vs RHE, where the lead-based catalysts exhibits a significantly lower extent of oxide formation, the onset of oxide formation was at much higher potential for the PtRu@Pb catalysts.

Cyclic voltammetric curves for methanol oxidation of the catalysts have been carried out 1M methanol and 0.1 M $HClO_4$ . Figure 3(b) shows the cyclic voltammogram of the anode electrocatalysts ( 60% PtRu –ETEK catalysts, PtRu@PbRT and PtRu@Pb 220(heat treated at 220°C) in Ar-saturated 0.1 M $HClO_4$ and 1M methanol solution. The methanol oxidation under identical condition showed that PtRu@Pb 220 catalysts had the highest current density in compare to commercial catalysts. To understand the kinetic pathways of the catalysts activity, the potential step experiments had been carried out in 1M methanol and 0.1 M $HClO_4$. The electrode surface was cleaned of pre-adsorped methanol decomposition products by stepping potential from a potential in the hydrogen region (0.085 V) to 0.855V for 10 s to oxidize all the preadsorbed products[32]. The transient currents were recorded from 0.3 to 0.9 V with a potential step increase of 50mV per experiment for time interval of 180 s. The steady-state current recorded at each potential is plotted against the potential and it clearly showed that the onset potential for methanol oxidation was practically same for the PtRu@Pb220, PtRu@PbRT and commercial ETEK catalysts (Figure 3c), in fact the chronoamperometric step experiment actually confirms that under identical situation, the methanol oxidation on PtRu@Pb220 surfaces showed much higher rate of methanol oxidation. When Semi-logarithmic log I vs electrode potential plots for methanol oxidations for all three catalysts had been plotted (Figure 3d), the Tafel slops for PtRu@Pb220, PtRu@PbRT was calculated to be 94-93 millivolts per dec whereas the Tafel slop for commercial ETEK catalysts was 91 millivolts per dec.

The long term chronoamperometric data for methanol oxidation had been collected for the PtRu@Pb220, PtRu@PbRT and commercial ETEK catalysts (Figure 4a) and indeed the PtRu@Pb220 showed the highest current density for long run. The cyclic voltammogram in 0.1M $HClO_4$ had been run for the PtRu@Pb220 catalyst to test the stability of the catalysts (Figure 4b). There was no noticeable reduction had been observed for electrochemical surface area of the PtRu@Pb catalysts. To understand the reason behind such a high current density of the PtRu@Pb catalysts, we have carried out the CO adsorption experiment where the electrode surface had been saturated with CO at a potential of 0.1 V in 0.1 M $HClO_4$ medium. The CO stripping experiment showed that the onset potential for CO stripping of PtRu@Pb 220 was much lower than the commercial PtRu-ETEK catalysts and the highest peak potential for CO stripping had been shifted by almost 60 millivolts (figure 4c). This clearly indicates may be the addition of lead on the PtRu alloy surfaces weakens the Pt-CO bonding. To understand the further mechanism of the activity of PtRu@Pb catalysts we have carried out the *in-situ* X-ray absorption spectroscopy in synchrotron facility.**EXAFS Analysis.** Traditional Fourier Transform (FT) EXAFS analysis was performed to explore the main structure of the catalysts. A *k*-range window of 2.74-10.86 $Å^{-1}$ (Hanning) and an *R*-window of 1.42-3.27 Å were used for all the EXAFS first shell fits. Due to the high correlation between the mean-square radial disorder $\sigma^2$ and the coordination numbers, the value of $\sigma^2$ was fixed to be the same for all fits to ensure the validity of the comparison of coordination numbers. The bond lengths $R_{Pt-Pt}$, and $R_{Pt--Ru}$, obtained by EXAFS analysis remained unchanged within uncertainties over different potentials, thus they were restricted to be constant over potentials to refine the fittings. This enabled us to perform the $\Delta\mu$ analysis that heavily relies on a crystallographic model with unchanging bond length.

Figure. 5b shows experimental data and theoretical fits in $R$ space at 0.8 V in 0.1 M $HClO_4$ electrolyte (without methanol). As seen, in the 2.5 Å region, the well-known region of highly destructive interference between Pt–Pt and Pt–Ru scattering was clearly observed evidence of alloying in the PtRu@Pb samples. A reasonably good EXAFS fit as revealed in figure 5b was achieved without including the Pt-Pb path. This was as expected because of the low concentration of lead in the alloy ($Pt:Pb_{Atomic\ ratio} \approx 1:3$) on the outer shell is beyond the sensitivity of EXAFS fitting. Due to the same reason, the bond distances between Pt-Pt and Pt-Ru obtained here, 2.72 Å and 2.67 Å, were not significantly different from those in PtRu(ETEK) alloy[33]. Therefore, while the compressive –strain effects induced by the shorter Pt-Pt distance is evidenced here, yet it does not play the dominate role in modification of the electronic properties of Pt on surface.

The variation of coordination numbers of Pt along with operation potentials in 0.1 M $HClO_4$ electrolytes is presented in figure5c. As shown, an abrupt increase in both Pt-Pt and Pt-Ru coordination numbers occurs at 0.8V. This widely observed phenomenon is due to the adsorption of OH at atop positions making the particles more spherical[34]. It shows that adsorption of O(H) from water activation becomes significant around 0.8 V, which removes all the remaining CO directly known as direct mechanism enabled by the electronic or ligand effects[35]. The direct evidence of O(H) adsorption on atop sites is given by $\Delta\mu$ signals as presented in figure6a. As shown, a prominent peak between 5-10 eV at 0.8V demonstrates the accumulation of O(H) onto Pt. In addition, the observed double peaks indicate the surface is dominated by two different metals (Pt and Ru in our case) because the shift of the OH/Pt peak close to M at the surface has been observed in PtM NPs only when M atoms are at the surface, which cause a core level shift of the Pt $L_3$ level nearby due to some charge transfer in between[35, 36]. Since the absolute $\Delta\mu$ amplitude is directly related to the adsorption coverage of that adsorbate[35], the height of the second peak in the highlighted region is evaluated to determine the relative amounts of O(H) present at different potentials.

The theoretical and experimental $\Delta\mu$ spectra of CO or O(H) adsorption on atop sites referred to clean surface are given in figure6b. As shown, the features of CO and O(H) theoretical $\Delta\mu$ spectra are significantly different in the energy range 0-5 eV relative to Pt $L_3$ edge: $\Delta\mu$(CO) and $\Delta\mu$(O(H)) give rise to negative and positive peak, respectively, allowing to distinguish the adsorption of these two species from each other. The negative peak observed in low potential region (< 0.6 V) experimentally identifies the adsorption of CO; on the other hand, the positive peak found in 0.8 V indicates the removal of CO and adsorption of O(H). The evolution of this peak from negative to positive along with the increase of potential clearly exhibits the direct mechanism: adsorption of O(H) on Pt removes the further CO, which is also observed for PtRu(ETEK)[33, 35]. Similarly, the absolute height of the peak in this highlighted region is evaluated to determine the relative amounts of CO or O(H) present at different potentials.

The variation of absolute $\Delta\mu$(CO) and $\Delta\mu$(O(H)) amplitudes along with potentials for PtRuPb was displayed in figure6c. For comparison, the corresponding $\Delta\mu$ amplitude collected on PtRu(ETEK) by Ramaker et al.[37] and $\Delta\mu$ amplitude for Pt/C with similar particle size are also included. As shown, at higher potential range (0.8 V or above), the O(H) coverage decrease in the direction PtRu > Pt > PtRu@Pb. The positive shift of the onset potential for PtRu@Pb compared to PtRu shown in figure3a affirms the stronger O(H) inhibition on PtRu@Pb than PtRu. Therefore, while alloying with Ru increases the binding energy between O(H) and Pt, the further deposition of Pb on surface significantly weakens the chemisorption of O(H), which makes the enhanced CO tolerance mechanism of PtRu@Pb completely different from that of PtRu. Specifically, for PtRu, a significant amount of CO survives from bifunctional mechanism[38, 39] (Eq. 2) due to the increased Pt-CO bond strength, which not only decelerates the following reaction rate, but heavily impedes CO migration across the surface to react with OH.

$$Pt-CO+Ru-OH \rightarrow Pt+Ru+CO_2+e^-+H^+ \qquad (2)$$

Afterwards, the remaining CO is continuously removed by the adsorption of O(H) on Pt near and far away from Ru[37]. Accordingly, Ramaker et al. claimed the ligand effects between Pt and Ru arise primarily from increased activation with water to form OH rather than weakening of the CO-Pt bond[35, 40]. As shown in Fig. 6c, the CO coverage on PtRu is still significant and reduces continuously at 0.4-0.8V along with the increase of O(H) coverage. On the other hand, the weaker O(H) chemisorptions on PtRu@Pb promotes the bifunctional mechanism by accelerating the reaction rate and facilitating the CO migration. As a result, much lower CO coverage with a smaller decreasing slope between 0.4-0.6V is observed (green solid line in Fig. 6c), and the sharper decrease beyond c.a. 0.65V is due to the adsorption of O(H) on Pt. Since bifunctional mechanism is widely acknowledged as the dominate effect in CO tolerance[41], we believe the better performance of PtRu@Pb can be mainly attributed to the weakening of the Pt-CO bond induced by the deposited Pb.

Using density-functional theory (DFT), Ranjan et al. found binding energies on $Pt_3Pb(111)$ are generally smaller than binding energies on Pt(111) (thus, weaker Pt-adsorbate interaction), mainly caused by the electron donation from Pb atoms to Pt atoms[25]. Experimentally, we observe lower white line intensity of Pt $L_3$ spectra in PtRu@Pb compared to Pt reference foil as shown in Fig. 7a (bottom). Since white line intensity is caused by the transition from 2p to unoccupied 5d orbitals, the reduction of it suggests the charge transfer toward Pt, filling the d-holes. On the contrary, PtRu nanoparticles usually show higher white line intensity than Pt foil[42]. In addition, as shown in Figure. 7a (top), the lower white intensity of PtRu@Pb is also observed by performing FEFF8 calculations on a $Pt_4Ru_2Pb$ cluster with Pb in the fcc adsorption site. The Pt-Pd bond distance is set to be 2.72 Å. Furthermore, the Pt *d-band* of PtRu@Pb was computed to be broader and lower in energy (Fig. 7b). The *d-band* center is calculated as the first moment of the projected density of *d* states (DOS) referred to the Fermi level, and the root mean squared (rms) *d-band* width was calculated as the second moment. According to the simple model developed by Hammer and Nørskov[43, 44], the downshift of the *d-band* center of a transition metal weakens the overall binding energy of simple adsorbates such as H, O, and CO. Thus, the role of deposited Pb in CO tolerance enhancement, which is promoting bifunctional mechanism by weakening the chemisorptions of O(H) through ligand effects, is suggested both experimentally and theoretically in this work.

It had been shown by a previously that the in situ infrared reflection absorption study of CO on underpotential-deposited Pb on Pt(111) in 0.1 M $HClO_4$ aqueous solution showed the reduction of the overall CO coverage to about a third than that observed at the same potential in the absence of Pb, preventing adsorption of CO on multi-bonded sites[45]. Similar trend had been observed in the case of our catalysts also, where the onset of CO stripping potential and the amount of CO coverage (calculated from the area under the curve) was found to be much less for PtRu@Pb220 catalysts in comparison to the commercial PtRu-ETEK catalysts. It is likely due to the faster desorption of CO from the PtRu@Pb220 surface. As mentioned earlier, the difference of work function between Pb and Pt is one of the important factor for the Pb-based Pt catalysts. Work function measurement proved that work function of Pt(111) electrode drastically decreases on deposition of lead[46]. The DFT-based calculation by other group indicates in contrast to Pt(111),[25] the increment of electron filling in the Pb-based intermetallic compounds results the rise of Fermi level, hence partial filling of 2π(CO)-d(Pt) antibonding level and thus weakens of Pt-CO bonding . Combining the result from the theoretical evidence with our in situ EXFAS study, we can conclude indeed lead play a significant role in the enhanced performance of PtRu@Pb catalysts.

Hence, we propose a model to explain the reaction mechanism operated by ternary catalysts which is as follows:

Step 1. $CH_3OH_{sol}$ + $Pt_{surfaces}$ of PtRu@Pb alloys → $CH_3OH_{ads}$ on Pt

Step 2. $CH_3OH_{ads}$ on Pt → $CO_{ads}$ on Pt (byproduct of methanol oxidation)

Step 3. $CO_{ads}$ on Pt + $Pb_{surfaces}$ of PtRu@Pb alloys → weakening of $CO_{ads}$ on Pt by Pb

Step 4. $H_2O$ + $Ru_{surfaces}$ of PtRu@Pb alloys → $OH_{ads}$ on Ru

Step 5. Weak $CO_{ads}$ on $Pt_{surfaces}$ + $OH_{ads}$ on Ru → $CO_2$ + $H^+$+$e^-$ + free Pt surfaces

The adsorption of methanol on Pt surfaces of the ternary alloys is the primary step of the mechanism, generating CO as a by-product of the methanol oxidation. We believe the strong bond formation between Pt-CO gets weaken by the charge transfer between lead (Pb) and Pt, resulting the faster removal of CO from the platinum surfaces and hence increases the adsorption of methanol on Pt surface. Though the contribution of hydroxyl group from ruthenium can not be overlooked, yet lead(Pb) plays more prominent role over Ru. Thus overall current density obtained is much high in comparison to the commercial PtRu catalysts. The earlier report [22] talked about a promoting effect of lead on the electro-oxidation of ethanol and it had been suggested that the primary mode of action of the lead ad atoms was the removal of adsorbed CO but there was no direct experimental evidence on that. Here we have shown by in situ XAS experiments, the exact role played by lead.

**Summery and Conclusion**

A novel ternary catalyst PtRu@Pb had been synthesized and microstructural and structural characterization of the catalysts confirmed the alloy phase formation with the incorporation of lead (Pb) in the crystal structure. The electrochemical studies have been performed using 0.1M $HClO_4$ and methanol oxidation study had been performed in a mixture of 0.1M $HClO_4$ and 1M methanol. The higher current density obtained by the methanol oxidation was many fold higher in comparison to the commercial PtRu catalysts

The CO stripping potential was found to be at lower potential than commercial PtRu catalysts which indicates that Pb influences the Pt-CO bond strength and helps to free active Pt surface sites faster and hence more methanol absorption could take place. That Results in the higher current density observed in the PtRu@Pb catalysts. The in situ X-ray Absorption Spectroscopy data suggested that decrease in the d-band vacancy of platinum on alloying with lead, weakens the overall binding energy of simple adsorbates such as H, O, and CO. Finally both the experimental and theoretical calculations suggest indeed lead helps for the better activity of the platinum in the ternary catalysts by changing the d-band state of Pt.

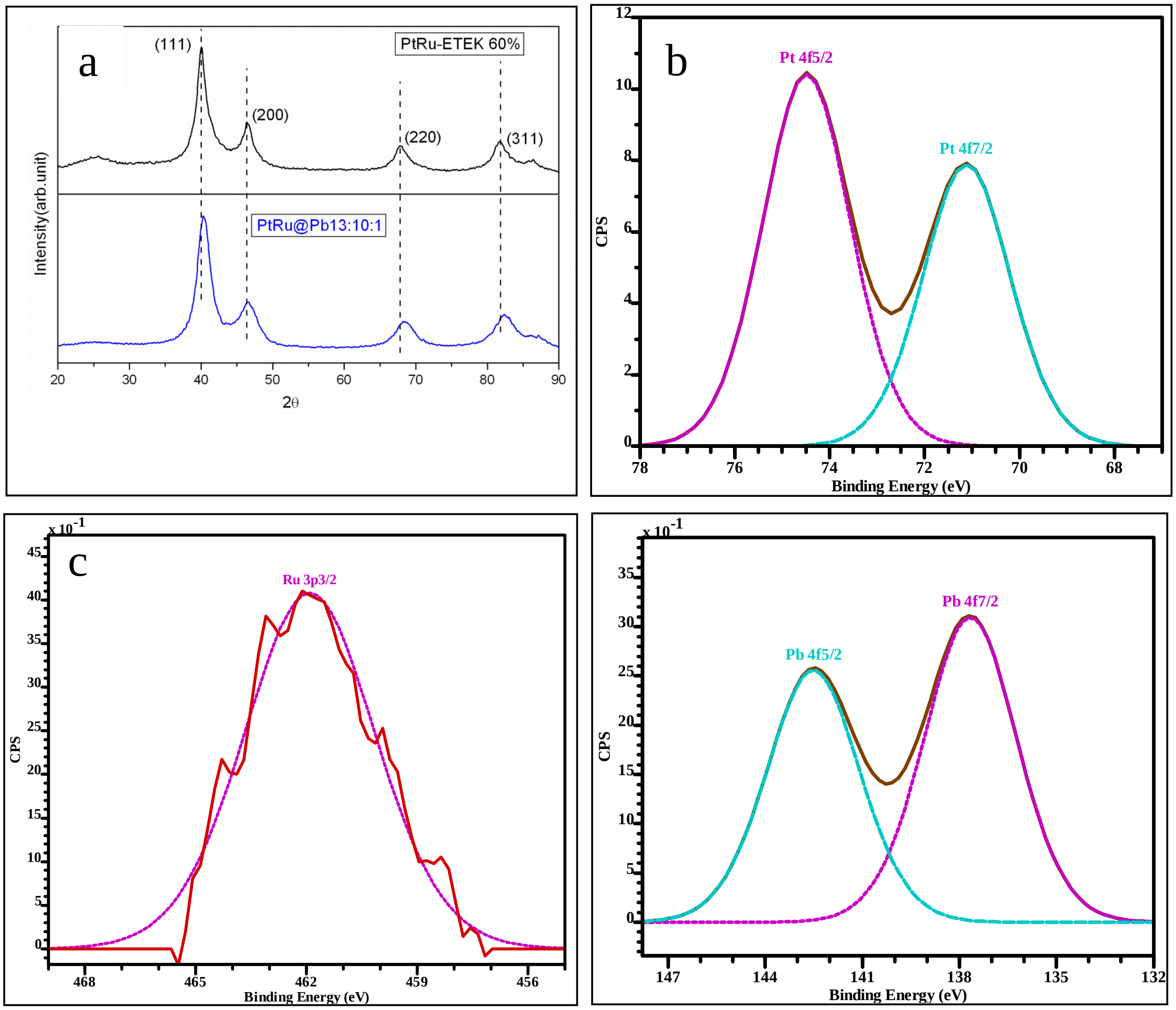

Figure1. (a )The XRD of the ternary catalysts and (b),(c) & (d) The X-ray Photoelectron Spectroscopy (XPS) spectra from Pt 4f , Ru 3p and Pb 4f respectively.

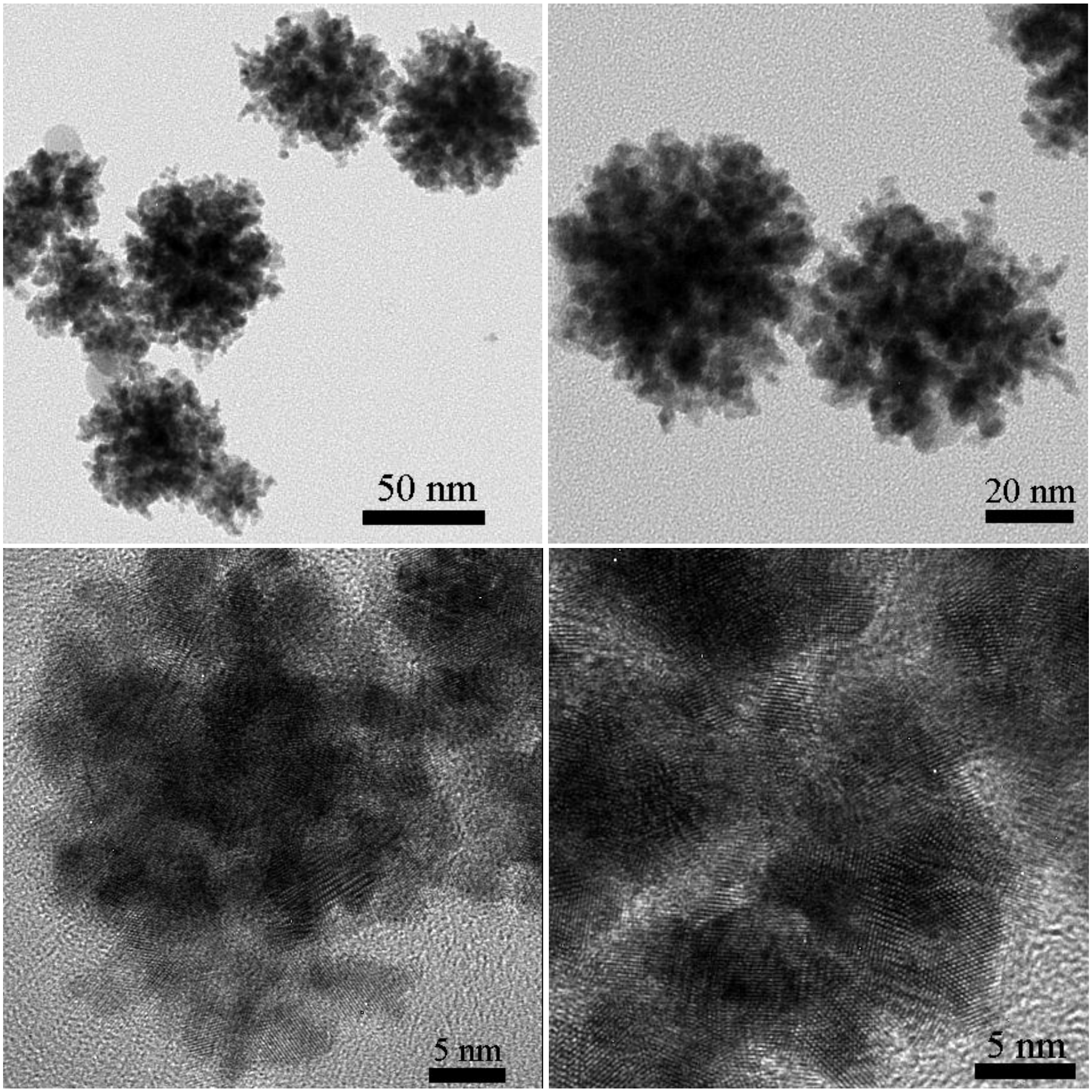

Figure2. The High resolution TEM micrographs of the ternary catalysts (a) shows the porous clusters of 50 nm of ternary PtRu@Pb catalysts supported on carbon, (b) reflects that each individual cluster is made of very small individual nanoparticles .(c) confirms that individual nanoparticles are  of 3-5 nm and (d) high resolution lattice fringes of the alloys and moire-fringes due to overlapping of nanocrystals of different lattice parameters.

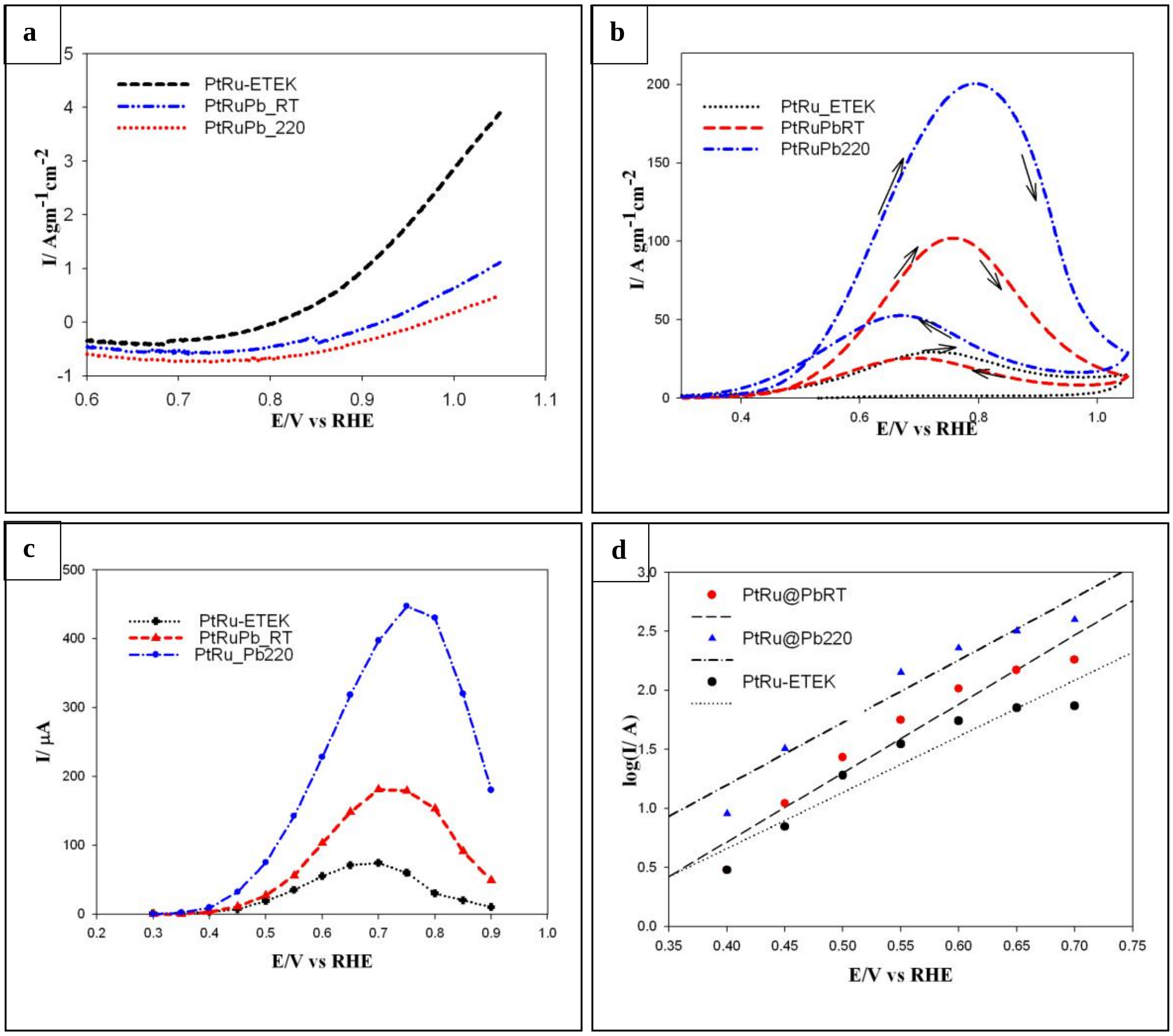

Figure 3. (a) The comparison of onset potential for the oxide formation for the thin films of 60% PtRu –ETEK catalysts, PtRu@Pb (umtreated/RT) and PtRu@Pb 220(heat treated at 220°C )catalysts in Ar-saturated 0.1 M $HClO_4$ solution. (b) Cyclic voltammogram of methanol oxidation for a thin film of 60% PtRu –ETEK catalysts, PtRu@Pb RT and PtRu@Pb 220 catalysts in Ar-saturated 0.1 M $HClO_4$ and 1M methanol solution.(c) Steady state current of the methanol oxidation for a thin film of 60% PtRu –ETEK catalysts, PtRu@Pb RT and PtRu@Pb 220 catalysts in Ar-saturated 0.1 M $HClO_4$ and 1M methanol solution. The steady state current was recorded from the chronoamperometric transients at t= 180 S. (d) Semilogarithmic logi vs electrode potential plots for methanol oxidations for a thin film of 60% PtRu –ETEK catalysts, PtRu@Pb RT and PtRu@Pb 220 in Ar-saturated 0.1 M $HClO_4$ and 1M methanol solution.*The loading of the catalysts material on the 5 mm diameter glassy-carbon electrode was 15µg/cm$^2$*.

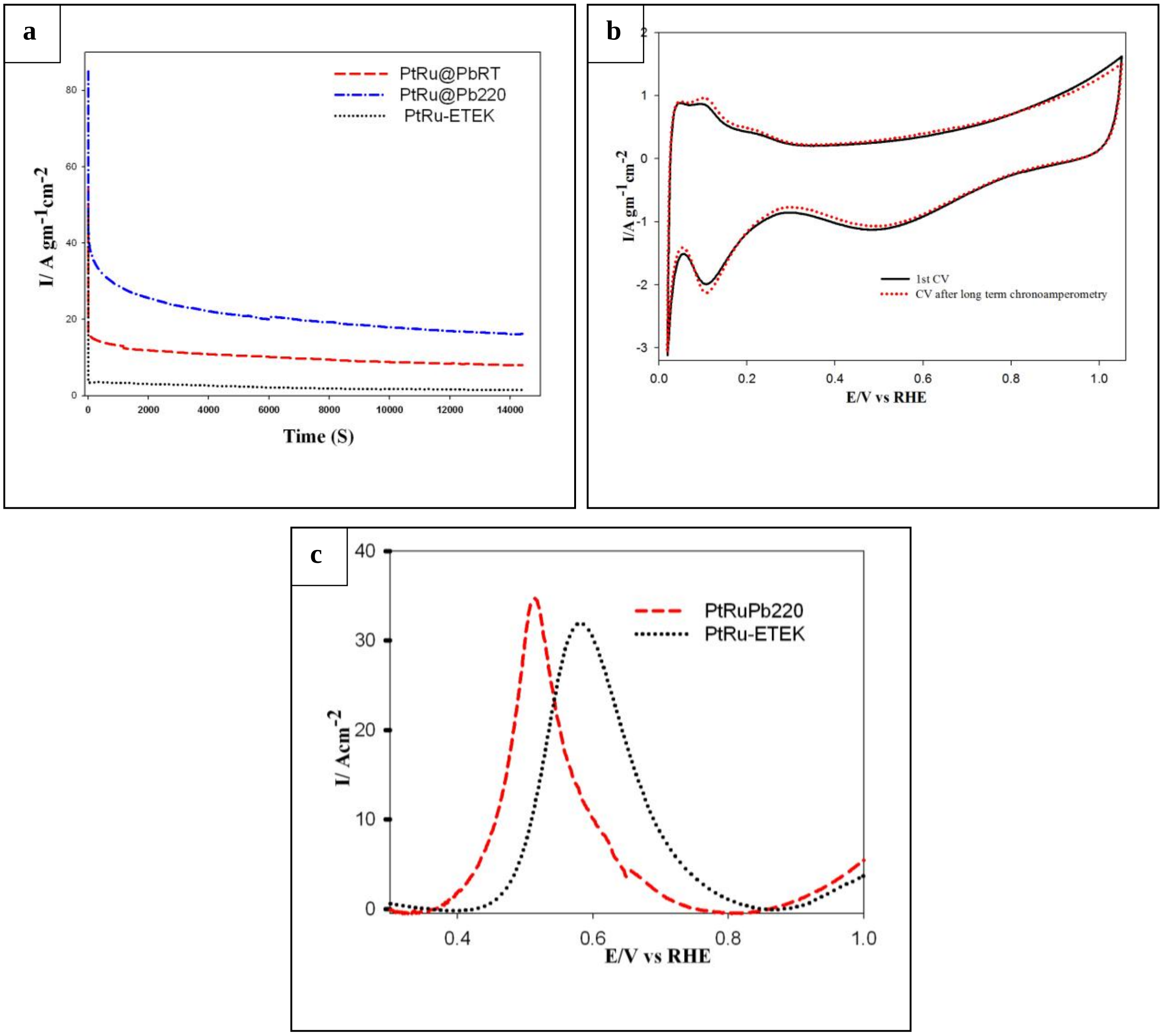

Figure 4. (a)Steady state long term chronoamperometry data of the methanol oxidation for a thin film of 60% PtRu –ETEK catalysts, PtRu@Pb RT and PtRu@Pb 220 catalysts in Ar-saturated 0.1 M $HClO_4$ and 1M methanol solution at 550mV. (b) The cyclic voltammogram of PtRu@Pb 220 catalysts in Ar-saturated 0.1 M $HClO_4$ solution after running long term (methanol oxidation) chronoamperometry at 550mV. (c) The CO oxidation by a thin film of 60% PtRu –ETEK catalysts, and PtRu@Pb 220 catalysts in Ar-saturated 0.1 M $HClO_4$ and 1M methanol solution. *The loading of the catalysts material on the 5 mm diameter glassy-carbon electrode was 15µg/cm$^2$.*

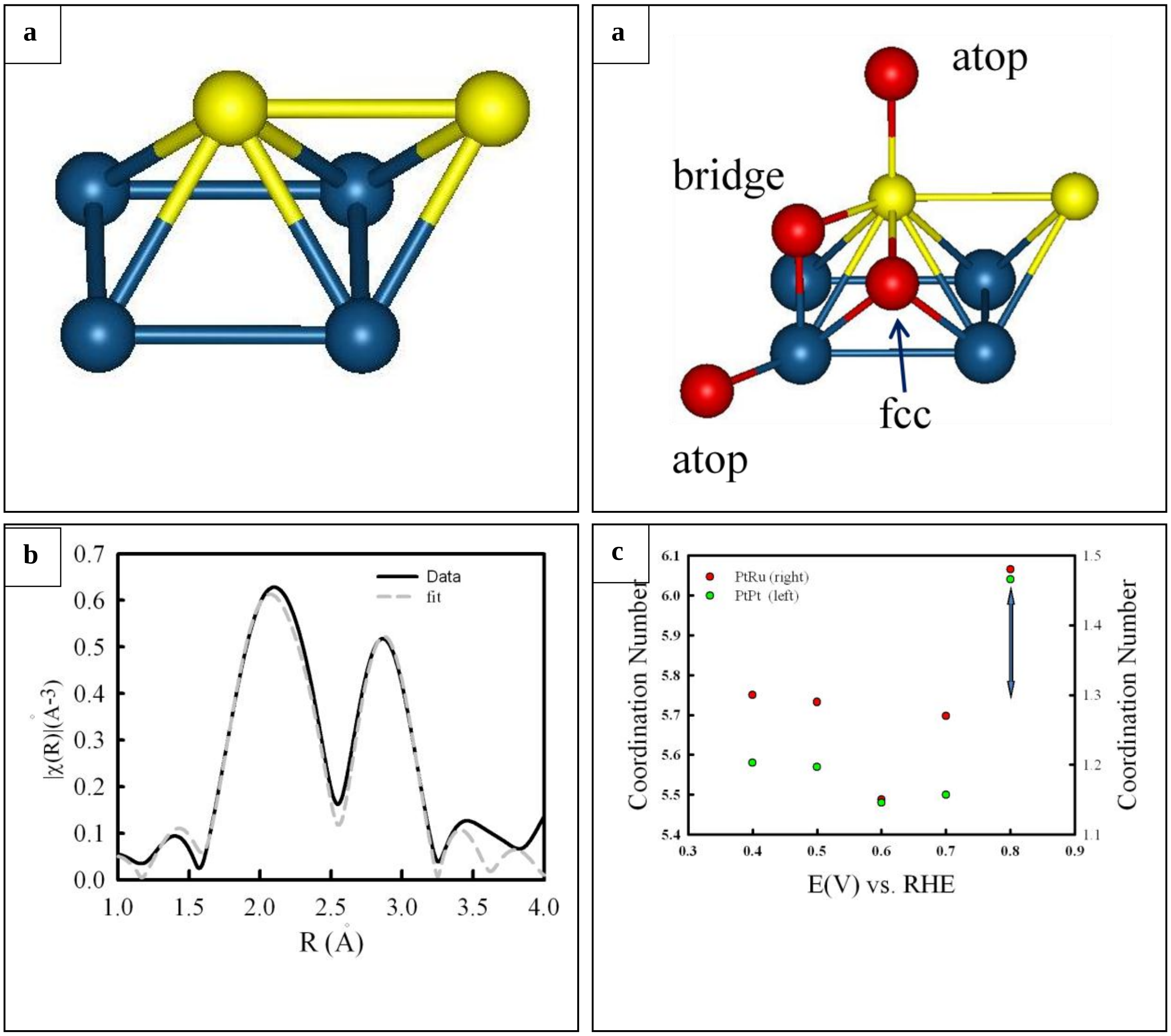

**Figure 5**. (a) $Pt_4Ru_2$ Janin cluster (left) for clean and (right) with adsorbate in atop, bridged, and fcc positions. In the Pt case, the Ru atoms are replaced by Pt. (b) Magnitude of FT for PtRu@Pb sample in 0.1 M $HClO_4$ electrolyte at 0.8V, a *k*-range window of 2.74-10.86 $Å^{-1}$ (Hanning) and a *R*-window of 1.42-3.27 Å were used for the fit. (c)Variation of coordination numbers of Pt (left) N_Pt, (right) N_Ru along with operation potentials.

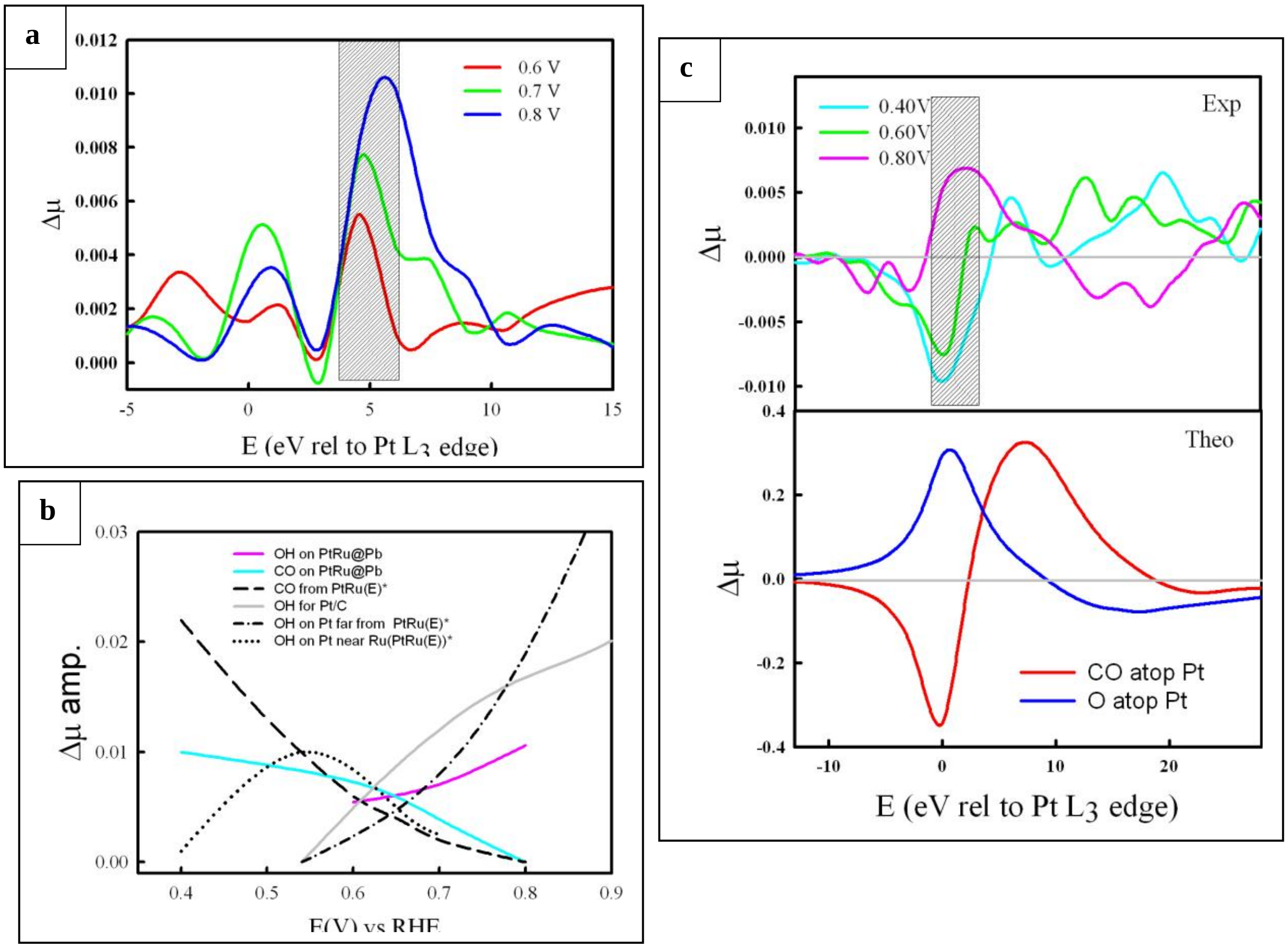

**Figure 6**. (a) Pt $L_3$ edge Δμ spectra for PtRu@Pb sample in clean 0.1 M $HClO_4$ using 0.50 V as the reference. The shaded areas indicate regions utilized to determine relative O(H) coverage. (b)Pt $L_3$ edge Δμ spectra for PtRu@Pb sample in 0.1 M $HClO_4$ containing 1 M methanol using 0.50 V in clean 0.1 M $HClO_4$ as the reference. The shaded areas indicate regions utilized to determine relative CO/O(H) coverage.( c) Δμ XANES amplitudes for the various adsorbates on the indicated catalysts. For PtRu@Pb, Δμ amplitudes were obtained by the height of the shaded area in Fig. 6a and 6b; for Pt/C, same method as above was used to collect Δμ amplitudes; for PtRu, Δμ amplitudes were taken from Ref [36,39].

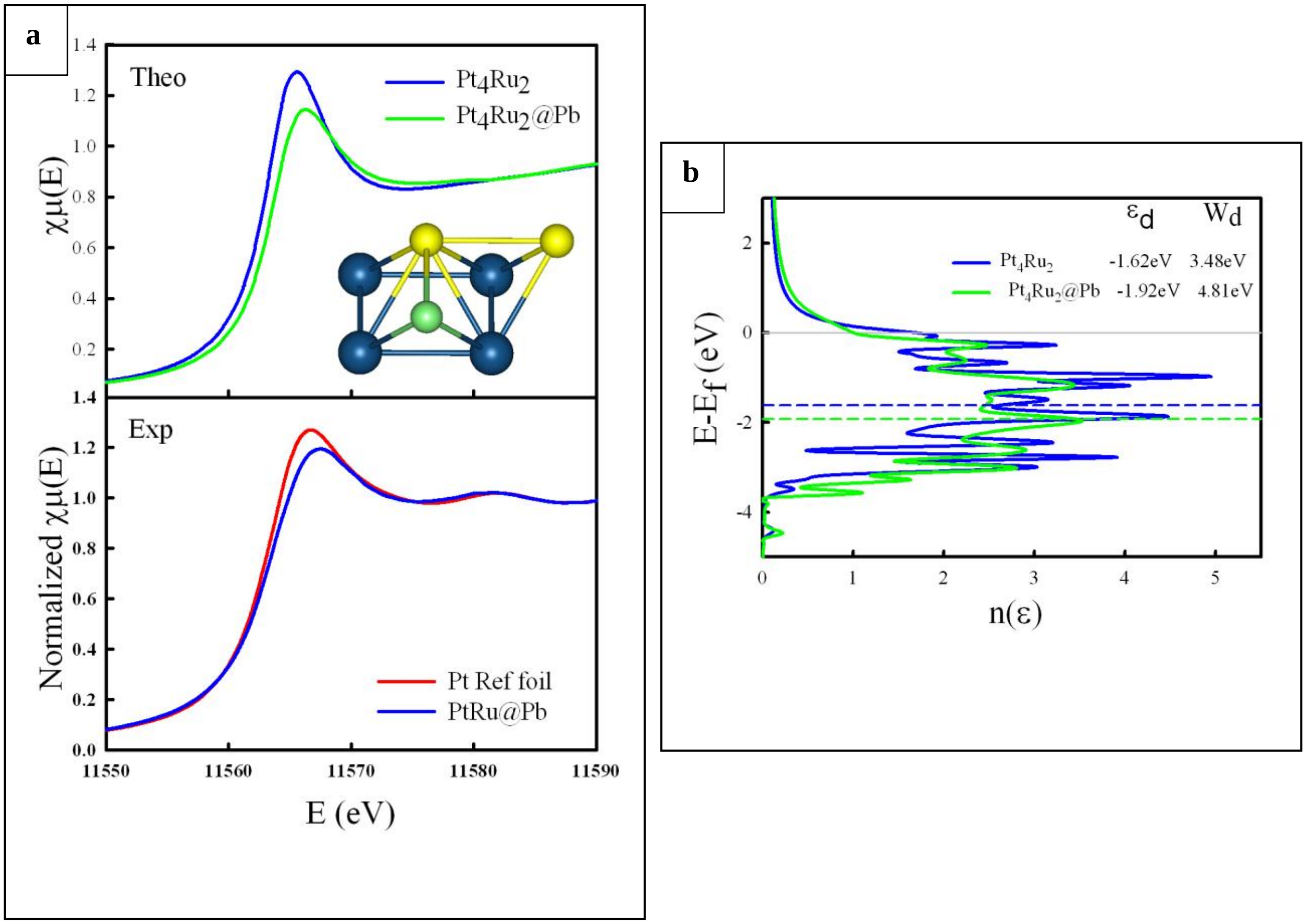

**Figure 7**. (a) Pt $L_3$ spectra for $Pt_4Ru_2$ and $Pt_4Ru_2$@Pb clusters calculated by FEFF8 program. A representative $Pt_4Ru_2$@Pb with Pb in fcc site is also shown (Top); experimental Pt $L_3$ spectra for of PtRu@Pb and the corresponding Pt reference foil (bottom). (b)The projected density of *d* states (DOS) referred to the Fermi level of $Pt_4Ru_2$ and $Pt_4Ru_2$@Pb clusters calculated by FEFF8 program. The *d-band* center ($\varepsilon_d$) is calculated as the first moment of DOS, and the root mean squared *d-band* width ($W_d$) was calculated as the second moment.